\documentstyle[12pt]{article}

\catcode`\@=11

\def\@normalsize{\@setsize\normalsize{15pt}\xiipt\@xiipt
\abovedisplayskip 14pt plus3pt minus3pt%
\belowdisplayskip \abovedisplayskip
\abovedisplayshortskip  \z@ plus3pt%
\belowdisplayshortskip  7pt plus3.5pt minus0pt}

\def\small{\@setsize\small{13.6pt}\xipt\@xipt
\abovedisplayskip 13pt plus3pt minus3pt%
\belowdisplayskip \abovedisplayskip
\abovedisplayshortskip  \z@ plus3pt%
\belowdisplayshortskip  7pt plus3.5pt minus0pt
\def\@listi{\parsep 4.5pt plus 2pt minus 1pt
            \itemsep \parsep
            \topsep 9pt plus 3pt minus 3pt}}

\def\underline#1{\relax\ifmmode\@@underline#1\else
        $\@@underline{\hbox{#1}}$\relax\fi}
\@twosidetrue

\relax

\catcode`@=12

\catcode`\@=11

\def\ps@headings{\def\@oddfoot{}\def\@evenfoot{}
\def\@oddhead{\hbox{}\hfill
        \makebox[.5\textwidth]{\raggedright\ignorespaces --\thepage{}--
\hfill {}}}}

\def\@evenhead{\@oddhead}
\ps@headings

\catcode`\@=12

\relax

\def\figcap{\section*{Figure Captions\markboth
        {FIGURECAPTIONS}{FIGURECAPTIONS}}\list
        {Fig. \arabic{enumi}:\hfill}{\settowidth\labelwidth{Fig. 999:}
        \leftmargin\labelwidth
        \advance\leftmargin\labelsep\usecounter{enumi}}}
 \relax
\def\tablecap{\section*{Table Captions\markboth
        {TABLECAPTIONS}{TABLECAPTIONS}}\list
        {Table \arabic{enumi}:\hfill}{\settowidth\labelwidth{Table 999:}
        \leftmargin\labelwidth
        \advance\leftmargin\labelsep\usecounter{enumi}}}
 \relax
\def\reflist{\section*{References\markboth
        {REFLIST}{REFLIST}}\list
        {[\arabic{enumi}]\hfill}{\settowidth\labelwidth{[999]}
        \leftmargin\labelwidth
        \advance\leftmargin\labelsep\usecounter{enumi}}}
 \relax

\catcode`\@=11

\def\@evenhead{\@oddhead}

\ps@headings

\relax

\newskip\humongous \humongous=0pt plus 1000pt minus 1000pt

\newif\ifdtup

\def\Tr{\mathop{\rm Tr}}

\def\beq{\begin{equation}}
\def\eeq{\end{equation}}

\def\beqn{\begin{eqnarray}}
\def\eeqn{\end{eqnarray}}
\relax

\def\G2{{\; \rm GeV/}c^2}
\def\G{\; \rm GeV}

\def\dotx{\dotx{\dot\overline{x}}}

\relax

\input {epsf}
\begin{document}

\begin{titlepage}
\nopagebreak
\begin{flushright}

        {\normalsize    OU-HET 190 \\
                     December,~1994  \\
                         revised version \\}

\end{flushright}

\vfill
\begin{center}
  {\large \bf  Continuum  Annulus Amplitudes  \\
            from the Two-Matrix Model }
\footnote{This work is supported in part by
  Grant-in-Aid for  Scientific Research
(06221245)~{\rm in~Priority~Area~and~}(05640347)
from
 the Ministry of Education, Japan.}

\vfill
           {\bf M. Anazawa}$~^{\dag}$,~~ {\bf A. Ishikawa}$~^{\ddag}$  \\
                          and \\
         {\bf H.~Itoyama}$~^{\dag}$  \\

\vfill
       $~^{\dag}$ Department of Physics,\\
        Faculty of Science, Osaka University,\\
        Toyonaka, Osaka, 560 Japan\\
                       and \\

       $~^{\ddag}$ Uji Research Center, Yukawa Institute for
 Theoretical Physics, \\
        Kyoto University, Uji, Kyoto, 611 Japan \\

\end{center}
\vfill


\begin{abstract}
An explicit expression for continuum annulus amplitudes having
 boundary lengths
$\ell_{1}$ and $\ell_{2}$ is obtained from the two-matrix model
for the case of the unitary series; $(p,q) = (m + 1, m)$.
 In the limit of vanishing cosmological constant, we find an integral
representation of these amplitudes which is reproduced,  for the cases of the
$m = 2~(c=0)$ and the $m \rightarrow \infty~(c=1)$,  by a continuum
approach consisting
of quantum mechanics of loops and a matter system integrated over
the modular parameter of the annulus.
 We comment on a possible relation to the unconventional branch of
the Liouville gravity.

\end{abstract}
\vfill
\end{titlepage}

 One of the intriguing properties of   the  noncritical strings with
$c \leq 1$
is  that the macroscopic $n$-loop amplitudes take a
suggestive form in terms of the  boundary lengths \cite{MSS}
which may inspire a
geometrical interpretation.  Properties of macroscopic loop amplitudes
encompass those of microscopic loop amplitudes from which we directly
extract the susceptibility and the operator dimensions of the continuum
theory.  They are, in principle, directly comparable with
 the results from the continuum path integrals on the geometry
  of annulus and the ones with more boundaries.
   The macroscopic loops may, in addition, represent
 the effects of boundary interactions of the theory.
 The derivation of the continuum loop (annulus) amplitudes has been given
in the one-matrix model at the multicritical
points both from the orthogonal polynomial approach \cite{MSS}( see also
\cite{BDSS}) and from the
Schwinger-Dyson approach \cite{AIMZ}.
No comparable work has been done, on the other hand,
for the case of the two-matrix model. (See \cite{MSS,MS}).
In this letter, we  report on a progress in this direction.


 The two-loop correlators we start with are
\beqn
W_{11} (\zeta_1, \zeta_2 ; \mu) &=& \langle \langle
                                     \Tr \frac{1}{X_1-\hat{M}}
                                     \Tr \frac{1}{X_2-\hat{M}}
                                   \rangle \rangle\;\;, \;\;\; \\
 W_{22} (\xi_1, \xi_2 ;\mu) &=& \langle \langle
                                     \Tr \frac{1}{Y_1-\hat{N}}
                                     \Tr \frac{1}{Y_2-\hat{N}}
                                   \rangle \rangle\;\;, \;\;\; \\
 W_{12} (\zeta, \xi ;\mu) &=& \langle \langle
                                     \Tr \frac{1}{X-\hat{M}}
                                     \Tr \frac{1}{Y-\hat{N}}
                                   \rangle \rangle\;\;.
\eeqn
Here, $\hat{M}$ and $\hat{N}$ are the matrix variables of the two-matrix
model. $X$'s and $Y$'s are the bare boundary cosmological constants and
$\zeta$'s and $\xi$'s are the renormalized boudary cosmological constants
in the sense of eq.~(\ref{Laplace}) below. We denote by $ \langle \langle
 \cdots \rangle \rangle $  the averaging in the planar limit.
Formulas have been obtained of these correlators  in \cite{DKK} for the
unitary cases $(p, q) = (m + 1, m),~~c = 1- \frac{6}{m(m+1)}$ by
 finding a parametrization
$ \zeta = \mu^m \cosh m \theta, ~\xi~ = \mu^m \cosh m \tau $;
\beqn
\mu \frac{\partial}{\partial \mu} W_{11} (\zeta_1, \zeta_2  ; \mu) &=&
  2\frac{\partial}{\partial \zeta_1} \frac{\partial}{\partial \zeta_2}
  \sum^{m-1}_{k=1} \frac{\sinh k \theta_1 }{\sinh m \theta_1}
                   \frac{\sinh k \theta_2 }{\sinh m \theta_2}\;\;,
 \;\; \label{formula}       \\
\mu \frac{\partial}{\partial \mu} W_{12} (\zeta, \xi ; \mu) &=&
 2\frac{\partial}{\partial \zeta} \frac{\partial}{\partial \xi}
  \sum^{m-1}_{k=1} (-)^{m-k} \frac{\sinh k \theta_1 }{\sinh m \theta_1}
                   \frac{\sinh k \theta_2 }{\sinh m \theta_2}\;\;.
  \label{formulanu}
\eeqn
We denote the renormalized cosmological constant   by $\mu^{2m} \equiv M^2$
   and $W_{11}, W_{12}$ and $W_{22}$  generically by $W$.
\footnote{  For the sake of simplicity and space, we have dealt with
  eq.~(\ref{formula}) explicitly in this letter.  Similar expressions
 can be derived from eq.~(\ref{formulanu}).}
 In what follows, we examine annulus amplitudes
$w(\ell_{1}, \ell_{2})_{c}$ having boundary
lengths $\ell_{1}$ and $\ell_{2}$.

$W(\zeta_1, \zeta_2  ; \mu)$ and $w(\ell_{1}, \ell_{2}; \mu)_{c}$ are related
by the Laplace transform
\beqn
\label{Laplace}
W(\zeta_1, \zeta_2  ; \mu) &=& \int^{\infty}_{0} d \ell_{1}
                             \int^{\infty}_{0} d \ell_{2}
                             e^{- \zeta_1 \ell_1} e^{- \zeta_2 \ell_2}
                             w(\ell_{1}, \ell_{2}; \mu)_{c}
\;\;\; \\ \nonumber
&\equiv& {\cal L}[w(\ell_{1}, \ell_{2})_{c}].
\eeqn
We have found the following formula for the inverse Laplace image
\beqn
\label{inverse}
{\cal L}^{-1}[\frac{\partial}{\partial \zeta}
         \frac{\sinh k \theta}{\sinh m \theta}]
= - \frac{M \ell}{\pi} \sin \frac{k \pi}{m}~  K_{\frac{k}{m}} (M \ell).
\eeqn
Note that $K_{\nu}(z)$ is the modified Bessel function.
{}From eqs.~(\ref{formula}),(\ref{Laplace}),(\ref{inverse}), we find
\beqn
\frac{\partial}{\partial M} w(\ell_1, \ell_2)_{c}
&=& {\cal L}^{-1}[\frac{\partial}{\partial M} W(\zeta_1, \zeta_2)]
\;\;\; \\ \nonumber
&=& \frac{2 M \ell_1 \ell_2}{m {\pi}^2} \sum^{m-1}_{k=1}
             (\sin \frac{k \pi}{m})^2
             K_{\frac{k}{m}}(M \ell_1) K_{\frac{k}{m}}(M \ell_2).
\eeqn
Integrating once, we obtain
\beqn
\label{result1}
w(\ell_1, \ell_2)_{c} = \frac{2}{m {\pi}^2}
                    \frac{M \ell_1 \ell_2}{\ell_1 + \ell_2}
                    \sum^{m-1}_{k=1}(\sin \frac{k \pi}{m})^2
                    K_{\frac{k}{m}}(M \ell_1) K_{1 - \frac{k}{m}}(M \ell_2).
\eeqn
  This is our main formula whose implications will be discussed below.
A similar but distinct formula is seen in \cite{Kos}.
An outline of the derivation of eq.~(\ref{inverse})
 as well as that of eq.~(\ref{result1}) will be given in the end
 of this letter.

\noindent
 One can easily check that eq.~(\ref{result1}) reproduces
the well-known two-loop amplitude for the case of
$(p, q) = (3, 2)$ ~\cite{MSS,Kos};
\beqn
w(\ell_1, \ell_2)_{c=0}
= \frac{1}{2 \pi} \frac{\sqrt{\ell_1 \ell_2}}{\ell_1 + \ell_2}
  e^{- M (\ell_1 + \ell_2)}.
\eeqn

Let us now take a close look at the limit of vanishing
cosmological constant $M \rightarrow 0$.
 From the asymptotic form of the product of two modified Bessel functions,
we find
\beqn
\label{asymptotics}
\lim_{M \rightarrow 0} M (\sin \frac{k \pi}{m})^2
         K_{\frac{k}{m}}(M \ell_1) K_{1 - \frac{k}{m}}(M \ell_2)
= (\frac{\pi}{2})^2
   \frac{1}{\Gamma(\frac{k}{m}) \Gamma(1 - \frac{k}{m})}
  (\frac{\ell_1}{2})^{- \frac{k}{m}}
  (\frac{\ell_2}{2})^{\frac{k}{m} - 1}.
\eeqn
We denote the two-loop amplitude in the limit of vanishing
cosmological constant by
\beqn
\label{definition}
w(\ell_1, \ell_2)^{M \rightarrow 0}_{c}
&\equiv& \lim_{M \rightarrow 0} w(\ell_1, \ell_2)_{c}.
\eeqn
{}From eq.~(\ref{result1}) and eq.~(\ref{asymptotics}), we find
\beqn
\label{M=0}
w(\ell_1, \ell_2)^{M \rightarrow 0}_{c}
&=& \frac{1}{m \pi} \frac{\ell_1}{\ell_1 + \ell_2}
    \sum^{m-1}_{k=1} \sin \frac{k \pi}{m}~~
          (\frac{\ell_2}{\ell_1})^{\frac{k}{m}}
\;\;\;  \nonumber  \\
&=& \frac{1}{m \pi} \sum^{\infty}_{n=0} \sum^{m-1}_{k=1}
           \sin (n + \frac{k}{m}) \pi ~~
          (\frac{\ell_2}{\ell_1})^{n + \frac{k}{m}}
\;\;\;, \; \;\; {\rm for}\;\;\; \ell_{1} > \ell_{2}.
\eeqn
In the Ising ($m = 3$) case, we can easily obtain a more explicit form
of the two-loop amplitude from  eq.~(\ref{M=0}),
\beqn
w(\ell_1, \ell_2)^{M \rightarrow 0}_{c = 1/2}
= \frac{1}{2 \sqrt{3} \pi}
\frac{(\ell_1 \ell_2)^{1/3}}{\ell_1 + \ell_2}
\left( (\ell_1)^{1/3} + (\ell_2)^{1/3} \right).
\eeqn
This is consistent with the result in \cite{MSS}.


 So far, we have found an explicit exression for the continuum
 annulus amplitudes
 (eq.~(\ref{result1}))  as well as the one
in the limit of vanishing cosmological constant (eq.~(\ref{M=0})).
We now study how this limit
may be reproduced by a  continuum framework. ( See also \cite{NKYM}
for a treatment at the proper-time gauge with the assumed weight
multiplicity). In this limit, the functional integral measure of the
matrix models concentrates on the boundaries; graphs are all degenerate and
the only interaction which would take place is at the boundaries.
We  first   point out an integral representation
for $w(\ell_{1}, \ell_{2})_{c}$ for the case of vanishing
 cosmological constant.

\beqn
\label{integral}
   w(\ell_{1}, \ell_{2})^{M \rightarrow 0}_{c}
&=& \frac{\sqrt{2 \beta}}{m \pi}
    \int^{\infty}_{0} \frac{dt}{t^{1/2}}
   e^{-\frac{\left( \log \ell_{2}/\ell_{1}\right)^{2}}
            {8 \pi \beta t}}
\;\;\; \\ \nonumber
&&\times
 \sum_{n=0}^{\infty} \sum^{m-1}_{k=1}
    (n + \frac{k}{m})
    \sin (n + \frac{k}{m}) \pi ~~
     e^{- 2 \pi \beta (n + \frac{k}{m})^2 t} \;\;\;,
\eeqn
where $\beta$ is an arbitrary parameter.

In the case of pure gravity ($m = 2$),
\beqn
\label{integral}
   w(\ell_{1}, \ell_{2})^{M \rightarrow 0}_{c=0}
&=& \frac{1}{4 \pi}
    \int^{\infty}_{0} \frac{dt}{t^{1/2}}
   e^{-\frac{\left( \log \ell_{2}/\ell_{1}\right)^{2}}
            {8 \pi t}}
\;\;\; \\ \nonumber
&&\times
 \sum_{n=0}^{\infty} (-)^n (2n + 1)
     e^{- 2 \pi (n + \frac{1}{2})^2 t} \;\;\;.
\eeqn
 The Jacobi triplet product identity
$ \eta(q)^{3} = [q^{1/24}
{\displaystyle \prod_{n=1}^{\infty} (1- q^{n}) ]^{3}
  = - \sum_{k = - \infty }^{\infty} (-)^{k}
 k q^{1/2 (k-1/2)^{2}}  } $
converts this expression into
\beqn
\label{eq:Jacobi}
  \frac{1}{4 \pi}  \int_{0}^{\infty}  \frac{dt}{t^{1/2}}
     e^{ - \frac{\left( \log \ell_{2}/\ell_{1} \right)^{2}}{8\pi t} }
 \eta( q=  e^{-4\pi t})^{3} \;\;.
\eeqn
Note that we have set $\beta = 1$ in this expression in order
 to make a comparison
 with the continuum result written in the standard notation.

 The appearance of the Dedekind function is  intriguing  and
we pause to discuss how  eq.~(\ref{eq:Jacobi}) is reproduced in
continuum  two-dimensional
gravity coupled to the conformal matter in the topology of annulus.
The standard treatment at the
conformal gauge yields the vacuum-to-vacuum  transition amplitude
\beqn
\label{eq:cont}
  {\cal Z}_{\rm cont} = \int_{0}^{\infty}  dt
 \left(  \frac{1}{\Omega ( CKV)}
  \frac{< \psi\mid \frac{\partial \hat{g}}{\partial t} >}{< \psi\mid
 \psi >^{1/2}}  ( {\det}^{\prime} P_{1}^{\dagger}P_{1} )^{1/2}
\right)_{\hat{g}}
    Z_{\phi} (t) Z_{m}(t) \;\;\;.
\eeqn
Note that the lengths $\ell_{1}$ and $\ell_{2}$ are not incorporated
in this formula. We have chosen the reference metric
$\hat{g} = \left( \begin{array}{cc}
   t^{2}& 0 \\ 0 &  1 \end{array} \right)$,
where $t$ is the modular papameter
of the annulus. $Z_{\phi} (t)$  is  the contribution from the path integral
with the Liouville action and  $Z_{m}(t)$
generically represents the contribution
from conformal matter fields.  The rest of the notations are standard and
we leave them
to  review articles. ( See, for instance, \cite{DP}.)
After  some calculation, we obtain
\beqn
 \frac{1}{\Omega( CKV)}  \frac{< \psi\mid \frac{\partial \hat{g}}
{\partial t} >}{< \psi\mid \psi >^{1/2}}  &=& \frac{\sqrt{2}}{t}\;\; \\
  \sqrt{2}   ( {\det}^{\prime} P_{1}^{\dagger}P_{1} )^{1/2} &=&
\left({\det}^{\prime}  \Delta_{ \hat{g}} \right)_{\rm Dirichlet} =
 \left({\det}^{\prime}   \Delta_{ \hat{g}}  \right)_{\rm Neumann}
  \;\;
  \nonumber  \\
  &=&  2t \eta( q= e^{-4\pi t} )^{2} \;\;,
\eeqn
which means that the parenthesis  $ \left( \cdots \right)_{\hat{g}}$
in eq.~(\ref{eq:cont}) is equal to $2 \eta( q= e^{-4\pi t} )^{2}$.
As for the $Z_{\phi}$,  the argument of \cite{DDK}
provides a translationally invariant flat measure $\left[ d \phi\right]$
and the Liouville
action which includes the zero mode $\phi_{0}(\sigma^{1})$.
  ( We expand  the Liouville field as $ \phi(\sigma^{1}, \sigma^{2})
 = \phi_{0} (\sigma^{1})  + {\displaystyle \sum_{ n \neq 0}}\phi_{n}
 (\sigma^{1}) e^{-2\pi i n \sigma^{2}}$ $ \equiv \phi_{0}
 (\sigma^{1}) + \tilde{\phi}
( \sigma^{1}, \sigma^{2})$.)
The difficulty of an attempt to find an
agreement between eq.~(\ref{eq:Jacobi}) and eq.~(\ref{eq:cont})
is that the zero mode part is not identifiable
as the length operator
$  \ell( \sigma^{1} ) \equiv  \int_{0}^{1}
 d \sigma^{2}  e^{\frac{\gamma}{2} \phi( \sigma^{1}, \sigma^{2}) }$.
This can be seen, for example, in
\beqn
\label{eq:eg}
  \dot{\phi}_{0} ( \sigma^{1}) = \frac{2}{\gamma} \frac{d}{d \sigma^{1}}
  \log \ell( \sigma^{1})
   -  \frac{ \int_{0}^{1} d \sigma^{2}
                       \dot{ \tilde{\phi}}
        ( \sigma^{1},\sigma^{2})
            e^{  \frac{\gamma}{2} \tilde{ \phi}( \sigma^{1}, \sigma^{2})
                                    }
               }
               {\int_{0}^{1} d \sigma^{2}
                e^{ \frac{\gamma}{2} \tilde{\phi}( \sigma^{1}, \sigma^{2})
                                    }         } \;\;\;.
\eeqn

  We propose to change the form of the action;
\beqn
  S_{\phi}^{\prime} = \frac{1}{ 2\pi \gamma^{2}} \int_{0}^{t} d \sigma^{1}
  \left( \frac{d}{d \sigma^{1}} \log \ell( \sigma^{1} ) \right)^{2}
  +  \frac{1}{ 8\pi} \sum_{n \neq 0} \int_{0}^{t} d \sigma^{1}
 \left( \dot{\phi_{n}} \dot{\phi}_{-n} + ( 2 \pi n )^{2} \phi_{n}\phi_{-n}
\right)
 \;\;\;.
\eeqn
 This is equivalent to ignoring the second part of eq.~(\ref{eq:eg}).
Accordingly
\beqn
 Z_{\phi} (t)  \rightarrow  Z_{\phi}^{\prime} (t)  =  \int [ {\cal D}
  \log \ell ] [{\cal D} \tilde{\phi}] e^{- S_{\phi}^{\prime}}
  =  \langle \log \ell_{2} ;t \mid \log \ell_{1} \rangle  \prod_{n \neq 0}
 \langle 0; t \mid 0 \rangle_{n}  \;\;\;.
\eeqn
 Here the first factor represents the transition amplitude
of quantum mechanics of the length variable and the remaining infinite
product represents the vacuum amplitude for an infinite number of
oscillators.
 Evaluating  this amplitude, we obtain
\beqn
\label{eq:contans}
  {\cal Z}_{\rm cont} \rightarrow
  {\cal Z}_{\rm cont}^{\prime}(\ell_{1}, \ell_{2}) =
  \frac{2 \sqrt{2}}{ \gamma} \int_{0}^{\infty} \frac{dt}{t^{1/2}}
  e^{- \frac{ \left( \log \ell_{2} /\ell_{1} \right)^{2}}
 { 2 \pi \gamma^{2} t} }  \eta ( q = e^{-4\pi t}) Z_{m}(t) \;\;.
\eeqn
 The factor $\frac{1}{t^{1/2}}$ comes  from
 $\langle \log \ell_{2} ;t \mid \log \ell_{1} \rangle$.
Comparing  eq.~(\ref{eq:contans}) with eq.~(\ref{eq:Jacobi}), we see
\beqn
\label{eq:zm}
  \gamma^{2} =4 \;\;, \;\;  Z_{m}(t) =   \eta( e^{-4 \pi t})^{2}
  \;\;.
\eeqn
 Although our discussion  leading to eq.~(\ref{eq:contans}) is heuristic,
let us take that the argument of \cite{DDK} can be applied here.
 This will give us $ \gamma = \frac{1}{2 \sqrt{3}}
\left( \sqrt{1-c} \mp \sqrt{25-c} \right)$ and to recover the correct
semi-classical limit in the spherical topology, we must
choose the minus sign in this formula.
 We find that $ \gamma^{2} =4$ is obtained if and only if $c=-2$  and
we choose the unconventional branch, {\it i.e.} the plus
sign.\footnote{It is curious that this unconventional branch also appears
in the recent discussion of touching interacions on random surfaces
\cite{Kleb}. }
The expression eq.~(\ref{eq:zm}) for the $Z_{m}(t)$
is reproduced by the path integral of the first order system
in which only the nonzero oscillating modes are included.
Note that our discussion essentially differs from that of \cite{Dist}  in the
treatment of zero modes. The bosonization  formula of \cite{FMS} does not
apply in our case.

On the other hand, in the case of $c \rightarrow 1$
($m \rightarrow \infty$), the two-loop amplitude can be expressed
as
\beqn
\label{result2}
w(\ell_1, \ell_2)^{M \rightarrow 0}_{c \rightarrow 1}
&\equiv& \lim_{m \rightarrow \infty} \lim_{M \rightarrow 0}
w(\ell_1, \ell_2) \;\;\;  \nonumber \\
&=&
\frac{\sqrt{2 \beta}}{\pi}
   \int^{\infty}_{0} \frac{d t}{t^{1/2}}
   e^{- \frac{(\log \ell_1 / \ell_2)^2}{8 \pi \beta t}}
\;\;\;  \nonumber \\
&&\times \sum^{\infty}_{n=0} \int^{1}_{0}
       d \nu (n + \nu) \sin \left\{ (n + \nu) \pi \right\}
       e^{-2 \pi \beta (n+\nu)^2 t}
\;\;\;  \nonumber \\
&=& \frac{1}{8 \beta \pi} \int^{\infty}_{0} \frac{d t}{t^{1/2}}
     e^{- \frac{(\log \ell_1 / \ell_2)^2}{8 \pi \beta t}}
    \frac{1}{t^{3/2}} e^{-\frac{\pi}{8 \beta t}}
\;\;\; \\
&=& \frac{1}{(\log \frac{\ell_2}{\ell_1})^2 + \pi^2}.
\eeqn
Note that in this case we  have no criterion to fix $\beta$.
That eq.~(\ref{result2}) contains no Dedekind function
 has a well-known interpretation  in the continuum framework.
When the cosmological constant is vanishing, the Liouville field
acts as an extra conformal matter field.
Therefore  the target space is two-dimensional,
  which does not allow a string to vibrate.
 The cancellation of the nonzero modes
can be explicitly seen in the continuum
 both for the case of torus \cite{BK}
and  for the case of annulus \cite{Ishikawa}.


 We have derived the explicit expression for $w(\ell_{1}, \ell_{2})_{c}$
  as well as the one at $M \rightarrow 0 $. Our integral representation
  in this limit has an interpretation  from the continuum path
 integrals for the cases of $m=2~(c=0)$ and $m= \infty~ (c=1)$.
 Our discussion
suggests that, for the case of pure gravity, the amplitude is essentially
controlled by the lowest critical point $(2,1)$, where the central charge
is $c=-2$ and the only operator of the theory is the
boundary operator \cite{MMS}.  For the case of $c=1$, our result is consistent
 with the continuum calculation.


 Finally, we present an outline of the derivation of
eq.~(\ref{inverse}) and that of eq.~(\ref{result1})
to the extent space permits.
\noindent
We first quote a formula of a definite integral
\beqn
\int^{\infty}_{0} e^{- a x} K_{\nu} (x)
= \frac{\pi}{\sin \nu \pi}
  \frac{\sinh [\nu \log (a + \sqrt{a^2 - 1})]}{\sqrt{a^2 - 1}},
\eeqn
where $|\rm{Re}~ \nu| < 1$, $\rm{Re}~ a > -1$.
By setting $\nu$, $x$ and $a$  to be $\frac{k}{m}$, $M \ell$
and $\cosh m \theta$ respectively, we obtain
\beqn
\int^{\infty}_{0} e^{- \zeta \ell} K_{\frac{k}{m}} (M \ell) M d \ell
&=& \frac{\pi}{\sin \frac{k \pi}{m}}
    \frac{\sinh k \theta}{\sinh m \theta}
\;\;\; \\ \nonumber
&\equiv& {\cal L} [ M K_{\frac{k}{m}} (M \ell)].
\eeqn
Taking a derivate with respect to $\zeta$, we find
\beqn
\frac{\pi}{\sin \frac{k \pi}{m}} \frac{\partial}{\partial \zeta}
    \frac{\sinh k \theta}{\sinh m \theta}
= - {\cal L} [M \ell K_{\frac{k}{m}} (M \ell)].
\eeqn
The inverse Laplace transform of this eq.  provides eq.~(\ref{inverse}).

We  quote another formula of an indefinite integral
\beqn
\int^{z} dz z Z_{\nu} (\alpha z) Z^{*}_{\nu} (\beta z)
= \frac{z}{\alpha^2 - \beta^2}
  \left(
        \beta Z_{\nu} (\alpha z) Z^{*}_{\nu - 1} (\beta z)
      - \alpha Z_{\nu - 1} (\alpha z) Z^{*}_{\nu} (\beta z)
  \right),
\eeqn
where $\alpha \not= \beta$ and $Z_{\nu}(z)$ and $Z^{*}_{\nu}(z)$ represent
 either the Bessel function, the Neumann function or the Hankel function.
If we take $Z_{\nu}(z)$ and $Z^{*}_{\nu}(z)$ to be the Hankel function
$H^{(1)}(z) = \frac{2}{i \pi} e^{-i \pi \nu / 2} K_{\nu} (- i z)$,
we find
\beqn
\int^{z} dz z K_{\nu} (\alpha z) K_{\nu} (\beta z)
= \frac{z}{\beta^2 - \alpha^2}
  \left(
        \beta K_{\nu} (\alpha z) K_{\nu - 1} (\beta z)
      - \alpha K_{\nu - 1} (\alpha z) K_{\nu} (\beta z)
  \right).
\eeqn
 Setting ($z$, $\nu$, $\alpha$, $\beta$) to be ($M$, $\frac{k}{m}$,
$\ell_1$, $\ell_2$) and  using the identity $K_{\nu} = K_{- \nu}$,
we obtain the  formula eq.~(\ref{result1}).


\end{document}